# Ion Distributions at the Nitrobenzene-Water Interface Electrified by a Common Ion

Guangming Luo[a], Sarka Malkova[a], Jaesung Yoon[a], David G. Schultz[a,b], Binhua Lin[c], Mati Meron[c], Ilan Benjamin[d], Petr Vanýsek[e*], and Mark L. Schlossman[a,b*]

[a]Department of Physics, University of Illinois at Chicago, Chicago, IL 60607 USA, schloss@uic.edu

[b]Department of Chemistry, University of Illinois at Chicago, Chicago, IL 60607 USA

[c]The Center for Advanced Radiation Sources, University of Chicago, Chicago, IL 60637 USA

[d]Department of Chemistry, University of California, Santa Cruz, CA , 95064 USA

[e]Department of Chemistry and Biochemistry, Northern Illinois University, DeKalb, IL 60115 USA  pvanysek@niu.edu

Corresponding author:  Mark Schlossman, Department of Physics, University of Illinois at Chicago, Chicago, IL 60607 USA, schloss@uic.edu, 1 312 996 8787; FAX: 1 312 996 9016

**Abstract**

Synchrotron x-ray reflectivity is used to study ion distributions at the liquid-liquid interface between a nitrobenzene solution of tetrabutylammonium tetraphenylborate (TBATPB) and a water solution of tetrabutylammonium bromide (TBABr).  The concentration of TBABr is varied to alter the ion distribution.  Our x-ray measurements are inconsistent with several commonly used theories of ion distributions, including Gouy-Chapman, modified Verwey-Niessen, and the MPB5 version of the Poisson-Boltzmann equation.   These structural measurements are described well by ion distributions predicted by a version of the Poisson-Boltzmann equation that explicitly includes a free energy profile for ion transfer across the interface when this profile is described by a simple analytic form or by a potential of mean force from molecular dynamics simulations.  These x-ray measurements from the liquid-liquid interface provide evidence for the importance of interfacial liquid structure in determining interfacial ion distributions.

## 1. Introduction

Ion distributions near charged interfaces in electrolyte solutions underlie many electrochemical and biological processes, including electron and ion transfer across interfaces. The theory of Gouy and Chapman predicts the ion distribution near a charged, planar surface by solving the Poisson-Boltzmann equation with simplifying assumptions that include point-like ions and continuum solvents with uniform dielectric properties [1-3]. This mean field approach ignores the molecular-scale structure of the liquid.

An extensive development of theory has addressed these assumptions [4-8]. The ion distributions near the interface are expected to deviate from the Gouy-Chapman predictions largely as a result of the difference between the interfacial and bulk liquid structure. Few experimental probes are directly sensitive to the structure immediately adjacent to the charged interface. As a result, the Gouy-Chapman theory has not been thoroughly tested, yet it is the description of the ionic distribution most commonly used to analyze experimental data. Validation of this theory in the immediate vicinity of the interface should not be neglected, since it is this region that has the most influence on electron or ion transfer processes at the interface.

Ion distributions are important for processes at a variety of interfaces, including processes at solid electrodes, charged biomembranes, biomolecules, mineral surfaces, and at the liquid-liquid interface between two electrolyte solutions that is the subject of this study. Liquid-liquid interfaces underlie many practical applications in analytical chemistry and electrochemistry, and serve as model systems for reaction kinetics and biomimetic systems [9-12]. An advantage of the liquid-liquid interface for the study of ion distributions is that it does not impose an external structure on the adjacent liquid as might be expected from the atomic scale corrugations on a solid surface.

The liquid-liquid interface is formed between an aqueous solution of hydrophilic ions and a polar organic solution of hydrophobic ions. The ions form back-to-back electrical double layers, as suggested by Verwey and Niessen [13]. The samples studied here contain a common ion soluble in both phases, though not to the same extent. Partitioning of the ions in the bulk phases produces an electric potential difference between the phases. By varying the initial solution concentration of the common ion in one of the phases, the potential difference and the structure of the electrical double layers can be varied.

In analogy with the interface between a solid electrode and a liquid, the early theories of the liquid-liquid interface treated it as sharp and flat. Verwey and Niessen treated the interface as consisting of two back-to-back Gouy-Chapman ion distributions (i.e., a double layer on both sides of the interface) [13]. Classical electrochemical measurements of the liquid-liquid interface have discovered inadequacies in this approach. Capacitance measurements as a function of applied bias potential at the liquid-liquid interface depend upon the ionic species [14-20] in contradiction with the Gouy-Chapman theory for which only the ionic charge is relevant. In addition, the shape and magnitude of the capacitance as a function of interfacial potential are often in disagreement with Gouy-Chapman theory [15, 19-21].

A variation of the Gouy-Chapman theory, the modified Verwey-Niessen model, was motivated by the Stern model that postulated an inner layer of solvent at the metal-solution interface [22, 23]. Similarly, the modified Verwey-Niessen model describes the liquid-liquid interface as consisting of a compact layer that separates two Gouy-Chapman ion distributions (or diffuse layers) [24, 25]. This was applied, with partial success, to explain capacitance measurements at the liquid-liquid interface [14-18, 26].

Schmickler *et al.* used a mixed boundary layer, similar to a van der Waals interface, to consider the effect on the capacitance of overlapping ion distributions from each phase [19, 20, 27]. Urbakh and co-authors allowed for ion penetration into the other phase (as in the mixed boundary layer approach) by using a free energy profile of ion transfer that varied smoothly through the interface [28, 29]. A permittivity function that varied smoothly through the interface was also employed. Urbakh and co-authors provided a general formalism for predicting the capacitance if the permittivity function and the free energy profile function are known. Unfortunately, these two functions are not generally known and direct experimental tests of them are not available. In this paper, we show how to use x-ray reflectivity to probe the free energy profile function.

It is well known that liquid interfaces are not mathematically flat, rather, they exhibit fluctuations driven by the available thermal energy. These capillary wave fluctuations were postulated in the 1960's [30]. A combination of light scattering and, more recently, x-ray scattering has probed these fluctuations over length scales from angstroms to micrometers at liquid-vapor and liquid-liquid interfaces [31-34]. Urbakh and co-authors studied theoretically the effect of capillary waves on the ion distributions. For large concentrations, they showed that the ion distributions will be distorted by the capillary waves [35]. A recent density functional theory of the liquid-vapor interface of van der Waals liquids indicates that the interfacial profile will be distorted for capillary wave wavelengths on the order of the molecular size [36]. Our analysis assumes that the local ion and solvent distributions are not distorted by the presence of capillary waves because we are primarily sensitive to capillary wavelengths longer than the molecular size and we use relatively low concentration solutions.

Interfacial fluctuations are visible in the results of molecular dynamics simulations of neat liquid-liquid interfaces [37-39]. These simulations reveal an interface that is locally sharp, but fluctuating with capillary waves. Our recent x-ray reflectivity measurements have confirmed this picture for the nitrobenzene-water and 2-heptanone/water interfaces [40-42]. Although it has been computationally intractable to carry out realistic molecular dynamics simulations of the liquid-liquid interface between solutions at electrolyte concentrations typical of the samples studied in this paper, it is possible to study a single ion at a neat liquid-liquid interface. By positioning the ion at different distances from the interface and calculating the force on the ion at each position, the potential of mean force can be determined [37, 43-45]. This is equivalent to the free energy profile of ion transfer for a single ion. We demonstrate how the potential of mean force can be used in the analysis of x-ray reflectivity data.

Few experimental tools can directly probe ion distributions in solutions near interfaces. The surface scattering of x-rays and neutrons are, in principle, sensitive to this distribution. In particular, several x-ray studies have explored the electrical double layer for different geometries. Bedzyk et al. used long period x-ray standing waves to study the double layer adjacent to a charged phospholipid monolayer adsorbed onto a solid surface [46]. Their data were consistent with an adsorbed Stern layer of ions and a diffuse layer described by the linearized Gouy-Chapman model that yields an exponentially decaying charge distribution.

A number of x-ray studies have probed the structure of the Stern layer (of ions) due to counterion adsorption to a Langmuir monolayer on the surface of water, but did not make conclusions about the diffuse (or Gouy-Chapman) part of the ionic distribution [47-50]. One of these studies suggested the presence of additional ions further from the surface than the Stern layer [49]. Fenter et al. used Bragg x-ray standing waves and surface x-ray absorption spectroscopy to probe the structure within the adsorbed (Stern) layer of an electrolyte solution on a mineral surface. They determined roughly the partitioning of the ions between the adsorbed layer and the diffuse charge layer, but could not probe the form of the ionic distribution [51]. Recent studies of counterion condensation around DNA using small angle x-ray scattering demonstrated agreement with the solutions of the non-linear Poisson-Boltzmann equation for this geometry that included an atomic model of the DNA [52]. Further studies by this group provided indirect evidence that ion size needs to be considered in the Poisson-Boltzmann treatment [53].

In this work x-ray reflectivity was used to study the nitrobenzene-water liquid-liquid interface electrified by the partitioning of a common ion, tetrabutylammonium, in the two phases. Samples with different concentrations of this common ion had different interfacial potentials and, subsequently, different ionic distributions. The x-rays probed the electron density profile determined by the ionic distributions. An earlier report of these data demonstrated large deviations from the predictions of Gouy-Chapman theory, except for the most dilute samples [54]. Here, we demonstrate that these data are also inconsistent with several other commonly used models of the liquid-liquid interface. By including an ion free energy profile in the Poisson-Boltzmann equation, in addition to the usual mean field dependence of ionic concentration on the local electric potential, the effect of ion sizes and ion-solvent correlations are included. The effect of ion-ion correlations was also considered, though they have a negligible effect on our samples. Our data agree with the predictions of this generalized Poisson-Boltzmann equation when the free energy profile is either the result of a molecular dynamics simulation (as reported earlier [54]) or a simple analytic form. We emphasize that the agreement between the experiment and the prediction based upon the molecular dynamics simulation does not require fitting of the x-ray data, and that there are no adjustable parameters.

## 2. Materials and Methods

**Materials** Bulk solutions of tetrabutylammonium bromide (TBABr from Fluka *puriss.*, electrochemical grade) in purified water (Milli-Q) and tetrabutylammonium tetraphenylborate (TBATPB from Fluka *puriss.*) in nitrobenzene (Fluka *puriss.* ≥99.5% filtered seven times through basic alumina) were placed into contact and equilibrated in glass bottles. The solutions were then placed into a glass beaker for tension measurements or into one of two types of x-ray sample cell. The liquids were kept at a room temperature of 24 ± 0.5°C.

**Sample Cells** One x-ray sample cell is vapor tight and fabricated from stainless steel. This cell had mylar x-ray windows and wall inserts arranged such that the liquid-liquid interface was in contact only with mylar. The sample cell design is similar to that described previously [55] except that the windows and wall inserts were slanted at an angle of 53° from the horizontal (Fig. 1). This removed most of the interfacial curvature as required for x-ray reflectivity measurements. Fine-tuning of the interfacial flatness was accomplished by rotation of the sample cell about an axis transverse to the x-ray beam, a variation of a previously published procedure [55]. This addressed the problem of curvature due to the contact angle of the interface at the side walls, which proved to be more problematic that the effect of the interfacial contact angle at the x-ray windows. The interfacial area was 75mm x 100mm (along the direction of the x-ray beam x transverse).

A second sample cell, used for some of the measurements, was fabricated from cylindrical glass with a 70 mm inside diameter and a 1 mm thick wall. The liquid-liquid interface was flattened by positioning the interface at the top of a teflon sheet wrapped inside the cylinder. The sheet was held in place with a strip of stainless steel shim stock that functioned as a cylindrical spring to press the teflon against the inner wall of the glass. The top edge of the teflon sheet pinned the interface, which was flattened by adjusting the volume of the lower phase of nitrobenzene. The cell was sealed on top with an o-ring. Small glass tubes extended from this cell (to allow the introduction of electrodes for other experiments to be reported elsewhere) and were sealed with rubber septa. The cell was reasonably leak tight. X-rays passed through the glass walls and through the upper, aqueous phase to reflect off the nitrobenzene-water interface.

**Solutions** The liquid solutions were prepared at a concentration of 0.01M TBATPB in nitrobenzene and concentrations of 0.01, 0.04, 0.05, 0.057, and 0.08M TBABr in water. Upon equilibration, the ions partitioned between the two phases until the electrochemical potential for each ion was equal in both phases. Use of a common ion, $TBA^+$, allowed the electric potential across the liquid-liquid interface to be varied by adjusting the concentration of TBABr at a fixed concentration of TBATPB. The ion partitioning and the interfacial electric potential $\Delta\phi^{w-nb} = \phi^w - \phi^{nb}$ (potential in the bulk water phase minus the potential in the bulk nitrobenzene phase) were calculated using the Nernst equation and the standard Gibbs energy of transfer of an ion from water to nitrobenzene, $\Delta G_{tr}^{o,w\to nb}$ (= –23.84 kJ/mol for $TBA^+$, –35.9 kJ/mol for $TPB^-$, and 31.06 kJ/mol for $Br^-$) [56-58]. The values for $Br^-$ and $TBA^+$ were taken from molecular dynamics simulations, but are similar to values in the literature [56, 58]. The bulk ion concentrations and interfacial potentials are listed in Table 1.

**Interfacial Tension** The interfacial tension was measured with a Cahn microbalance that measures the weight of a teflon Wilhelmy plate fully submerged in the top, water phase. The bottom edge of the plate was placed in contact with the liquid-liquid interface. We have used this method previously [40, 42, 59] to obtain tension values in excellent agreement with literature measurements that used the ring, pendant drop, and maximum bubble pressure methods [60-62]. Values of the tension for the samples studied are listed in Table 2 and are comparable to literature measurements [63].

**X-ray Methods** X-ray reflectivity was measured at the ChemMatCARS beamline 15-ID at the Advanced Photon Source (Argonne National Laboratory, USA) with a liquid surface instrument and measurement techniques described in detail elsewhere [55, 64-66]. The reflectivity data were measured as a function of the wave vector transfer normal to the interface, $Q_z = k_{scat} - k_{in} = (4\pi/\lambda)\sin\alpha$ (the in-plane wave vector components $Q_x = Q_y = 0$ where $\lambda = 0.41360 \pm 0.00005$ Å is the x-ray wavelength and $\alpha$ is the angle of reflection) (Fig. 1). The x-rays penetrated through the upper bulk water solution then reflected off the water-nitrobenzene interface. The absorption length for water at our x-ray wavelength is 31 mm. The reflectivity data consist of measurements of the x-ray intensity reflected from the sample interface normalized to the incident intensity measured just before the x-rays entered the sample. These measurements were taken through the specular ridge in $Q$-space by scanning the detector vertically, but orienting it so that it always points towards the sample center. This produced a peak with a flat top whose wings represent background scattering that arises primarily from the bulk liquids. A linear fit to the wings allowed for subtraction of the background. The remaining intensity in the peak is the reflected intensity for a given $Q_z$ [55]. The data were then divided by the intensity of the x-ray beam transmitted straight through the bulk water phase, itself normalized to the incident intensity, to produce the reflectivity $R(Q_z)$.

The incident beam cross section was set by a slit (typically 12$\mu$m x 3 mm, v x h) placed 68 cm before the sample. A gas ionization chamber before the sample measured the incident x-ray flux used to normalize the reflected intensity. Copper absorber foils placed between the slit and the ionization chamber were used to adjust the intensity of the incident beam for the lowest values of $Q_z$ (< 0.04 Å$^{-1}$). The sample was followed by a slit with a vertical gap of ~0.3 mm to reduce the background scattering, and a scintillator detector was preceded by a slit with a vertical gap of ~0.3 mm that set the detector resolution. The sample to detector slit distance was 68.6 cm and the slits were separated by 34.3 cm.

The measurements reported here were taken during two different trips to the synchrotron, using the two different sample cells previously described. New solutions were prepared for different concentrations. Measurements on each sample were repeated to test for stability and for radiation damage. After the initial full data set was measured, repeated measurements at several different values of $Q_z$ were carried out over a period of several (6 to 12) hours. No radiation damage was evident.

**Molecular Dynamics Simulations** The potential of mean force (PMF) for ion transfer across the interface between two immiscible liquids has been the subject of numerous studies utilizing different methods [37, 43-45]. We chose to use the integral of the average force acting on the ion center of mass: $\Delta A = A_2 - A_1 = -\int_{z_1}^{z_2} \langle F_z(z) \rangle dz$, where $F_z$ is the projection along the *z*-axis (normal to the interface) of the total force acting on the ion's center of mass.

The calculations were carried out at 298 K using the velocity version of the Verlet algorithm with a 1 fs integration time step [43]. The ion center of mass was held fixed at 100 locations (separated by 0.3 Å for Br$^-$ and 0.6 Å for TBA$^+$) along the interface normal and a 100 ps simulation at each fixed location was used to compute the average force.

The potential energy functions for water and nitrobenzene have been described in an earlier publication [38]. The flexible simple point charge model was used for water and an empirically derived intermolecular (all atoms) flexible model for nitrobenzene [38]. The potential energy for TBA$^+$ is based on a united atom description of the CH$_2$ and CH$_3$ groups. The intramolecular potential utilized the Amber 86 parameter set [67]. The ion-liquids interactions were based upon Lennard-Jones plus Coulomb potentials, where the partial charges on the atoms were obtained from *ab-initio* calculations [68].

## 3. Data and Analysis

Figure 2 illustrates the reflectivity data for all concentrations studied. Below the critical wave vector for total reflection $Q_c$ ($\approx$ 0.007Å$^{-1}$) the reflectivity is predicted to be close to unity, though we were unable to measure reflectivity at or below this small value of the wave vector. Above $Q_c$ the reflectivity falls rapidly until reaching a value of ~$10^{-8}$, a practical limit imposed by the background scattering from the bulk liquids. The reflectivity is progressively reduced with increasing sample concentration.

The structure of the liquid-liquid interface will be determined by the distribution of ions and solvent molecules, including the effect of capillary wave fluctuations of the interface. X-ray reflectivity probes the electron density profile of this distribution, where the profile $\langle \rho(z) \rangle_{xy}$ is the electron density as a function of interfacial depth (along the *z*-axis) that is averaged over the region of the x-ray footprint that lies in *x-y* plane of the interface.

In this section, data analysis using several different models for the electron density profiles will be discussed. These models were used to predict ion distributions from which the reflectivity was computed by the Parratt formalism [69]. Most of the models do not self-consistently account for the interfacial capillary wave fluctuations, which must be added as described below. Most of the data analysis was carried out by directly comparing the reflectivity calculated from these models to the measurements. In the case of the analytic model for the potential of mean force, the reflectivity data were fit to determine parameters in the model.

### 3.1 Error Function Interface Model

A step-function interfacial profile describes an interface without structure. It consists of a constant electron density in each bulk phase with an abrupt step-function crossover at the interface (at $z = 0$). This profile can occur only if thermal capillary wave fluctuations of the interface are absent. The effect of capillary waves on this step-function profile is to produce an electron density profile $\langle \rho(z) \rangle_{xy}$ that varies as an error function (erf), such that [70]

$$\langle \rho(z) \rangle_{xy} = \frac{1}{2}(\rho_w + \rho_{nb}) + \frac{1}{2}(\rho_w - \rho_{nb}) \, erf[z/\sigma\sqrt{2}] \text{ with } erf(z) = \frac{2}{\sqrt{\pi}} \int_0^z e^{-t^2} dt. \quad (1)$$

In this case, the reflectivity has a simple form, given by [71]:

$$R(Q_z) \approx \left| \frac{Q_z - Q_z^T}{Q_z + Q_z^T} \right|^2 \exp(-Q_z Q_z^T \sigma^2) \quad (2)$$

where $Q_z^T = (Q_z^2 - Q_c^2)^{1/2}$ is the z-component of the wave vector transfer with respect to the lower phase, and $Q_c \approx 4\sqrt{\Delta\rho \, r_e \, \pi}$ is the critical value of $Q_z$ for total reflection (the electron density difference between the bulk water and nitrobenzene phases $\Delta\rho = \rho_{nb} - \rho_w$, and $r_e \approx$ 2.818 fm). Eq. (2) has been previously used to fit x-ray reflectivity data from pure liquid-liquid interfaces [40, 42, 59, 66]. The interfacial width (or roughness) $\sigma$, due to the capillary waves, can be calculated using [72-74]

$$\sigma_{cap}^2 = \frac{k_B T}{2\pi} \int_{q_{min}}^{q_{max}} \frac{q \, dq}{\gamma q^2 + \Delta\rho_m g} \approx \frac{k_B T}{2\pi\gamma} \log\left(\frac{q_{max}}{q_{min}}\right) \quad (3)$$

where $k_B T$ is Boltzmann's constant times the temperature, $\gamma$ is the measured interfacial tension, $\Delta\rho_m$ is the mass density difference of the two phases, $g$ is the gravitational acceleration, and the term $\Delta\rho_m g \ll \gamma q_{min}^2$. The variable $q$ is the in-plane wave vector of the capillary waves. The limit $q_{min} = (2\pi/\lambda) \Delta\beta \sin\alpha$ is determined by the instrumental resolution that sets the largest in-plane capillary wavelength probed by the x-rays (the angular acceptance of the detector $\Delta\beta = 4.7\times10^{-4}$ rads) [59, 73, 74]. The limit, $q_{max}$, is determined by the cutoff for the smallest wavelength capillary waves that the interface can support. We have chosen $q_{max} = 2\pi/5$ Å$^{-1}$ where 5 Å is an approximate molecular size. Note that $\sigma_{cap}$ depends on $q_{max}$ logarithmically and is not very sensitive to its value [55].

The data in Fig. 2 can be fit by Eq. (2) to yield values of the interfacial width $\sigma$ (fits not shown, though they are essentially indistinguishable from the lines in Fig. 2). Table 2 lists the fit values $\sigma$, the values predicted by capillary wave theory using Eq.(3), and the measured values of the interfacial tension in Table 2. The fit values are larger than the capillary wave values, indicating the presence of additional structure at the interfaces as expected for electrolyte solutions. Interfacial ion adsorption that increases with concentration is indicated by the reduction in interfacial tension with increasing concentration, as predicted by the Gibbs adsorption equation.

In analyzing these data with models that specify the additional structure due to ion adsorption, we adopt the viewpoint of the hybrid capillary wave theory in which an intrinsic electron density profile is roughened by capillary waves [75, 76]. The intrinsic profile is due to the arrangement of ions and solvent molecules near the interface but does not include interfacial thermal fluctuations. In the analysis of x-ray reflectivity data, the intrinsic profile is often treated phenomenologically by expressing the interfacial width as $\sigma^2 = \sigma_o^2 + \sigma_{cap}^2$, where a single number $\sigma_o$ represents the intrinsic width [32, 42]. Here, we do not follow this procedure because we are interested in the arrangement of ions and solvent molecules near the interface. In practice we will employ a theory, such as the Gouy-Chapman theory that predicts ion distributions for a flat, planar interface, then we will roughen these distributions with capillary waves using a method described below.

### 3.2 Gouy-Chapman Analysis

The Poisson-Boltzmann equation is used to describe ion distributions near electrified interfaces:

$$\frac{d}{dz}\left(\varepsilon(z)\frac{d}{dz}\phi(z)\right) = -\sum_i e_i c_i^o \exp[-\Delta E_i(z)/k_B T] \qquad (4)$$

where $\phi(z)$ is the electric potential at a distance $z$ normal to the interface, $\varepsilon(z)$ is the permittivity function, $e_i$ and $c_i^o$ are the charge and bulk concentration of ion $i$, $\Delta E_i(z)$ is the energy of ion $i$ relative to its value in the bulk phase, and $k_B T$ is Boltzmann's constant times the temperature. The Gouy-Chapman theory of ion distributions assumes that $E_i$ depends only upon the electrostatic energy, such that $E_i(z) = e_i \phi(z)$, as well as assuming that the permittivity function is given by the constant bulk value all the way up to the interface, $\varepsilon(z) \equiv \varepsilon$ (for pure water and nitrobenzene the bulk relative permittivities are $\varepsilon_{r,water} = 78.54$ and $\varepsilon_{r,nitrobenzene} = 34.8$). The analytic solution to Eq.(4) for a planar geometry when $E_i = e_i \phi(z)$, and $\varepsilon(z) \equiv \varepsilon$ is the Gouy-Chapman theory [1-3, 77].

The distribution of ion $i$ along the interfacial normal is given by

$$c_i(z) = c_i^o \exp[-\Delta E_i(z)/k_B T]. \tag{5}$$

This is illustrated for the Gouy-Chapman theory for the concentrations of our samples in Figure 3. The ion distributions take the form of back-to-back double layers on either side of a sharp planar interface. On the water side of the interface the double layer consists of Br$^-$ and TBA$^+$ ions, with the Br$^-$ ion concentration enhanced over its bulk value and the TBA$^+$ ion concentration reduced from its bulk value. On the nitrobenzene side of the interface the double layer consists of Br$^-$, TPB$^-$, and TBA$^+$ ions, with the TBA$^+$ ion concentration enhanced over its bulk value and the Br$^-$ and TPB$^-$ ion concentrations both reduced from their bulk values. The maximum concentration of the ions at the interface increases with the concentration of TBABr in water.

The intrinsic electron density profile,

$$\rho_{intrinsic}(z) = \rho_{solvent} + \sum_i c_i(z)(N_i - v_i \rho_{solvent}) \tag{6}$$

is calculated from the ion distributions $c_i(z)$ given by Eq. (5) using the number of electrons $N_i$ for ion $i$, the ion volume $v_i$ in the solution, and the electron density of the solvent $\rho_{solvent}$. The ions were modeled as spheres of diameter 3.7 Å for Br$^-$, 8.6 Å for TBA$^+$, and 9.5 Å for TPB⁻ [78, 79]. The diameters of Br⁻ and TBA⁺ were determined from our molecular dynamics simulations of the radial distribution functions within solution and are consistent with literature values [78, 80, 81]. Calculation of the ion distribution by Eqs. (4) and (5) considered the charge to be located at the center of the sphere assigned to each ion. A more realistic distribution of the charge within the ion, including dipole and quadrupole effects, had a negligible effect on the ion distribution calculated from Eqs. (4) and (5). Calculation of the electron density profile in Eq. (6) involved distributing the ion charge throughout the volume of the ion with a Gaussian function. The electron density profile in Eq. (6) is referred to as an intrinsic profile because it does not include the effect of capillary waves.

The electron density profile $\langle \rho(z) \rangle_{xy}$ that includes the effect of capillary waves can be calculated by convoluting the intrinsic electron density profile in Eq. (6) with a Gaussian of width $\sigma_{cap}$,

$$\langle \rho(z) \rangle_{xy} = \frac{1}{\sigma_{cap}\sqrt{2\pi}} \int_{-\infty}^{\infty} \rho_{intrinsic}(z') \exp[-(z-z')^2 / 2\sigma_{cap}^2] \, dz' \, . \tag{7}$$

The interfacial width $\sigma_{cap}$ is calculated using capillary wave theory from the measured interfacial tension (Table 2), as described previously [59]. The electron density $\langle \rho(z) \rangle_{xy}$ calculated from the Gouy-Chapman model for all samples is shown in Figure 4. Although the ion distribution in Figure 3 is discontinuous, the electron density profile is continuous because of the presence of capillary waves.

The Parratt algorithm was used to calculate the x-ray reflectivity by dividing the electron density profile into many layers, and determining the reflection and transmission coefficients at each of the interfaces between the layers [69]. Several hundred layers were used (with a layer spacing of 0.2Å in the interfacial region) to model the continuously varying density. The reflectivities calculated from the Gouy-Chapman model are shown as the dashed lines in Figure 5. At the lowest concentration this model matches the reflectivity data, but at higher concentrations the calculated reflectivities from the model differ by many standard deviations from the measured data. This indicates a failure of the Gouy-Chapman model. The Gouy-Chapman model overestimates the ion adsorption because it does not account for interfacial liquid structure due to the ion and solvent sizes and correlations between ions and solvent molecules.

### 3.3 Potential of Mean Force Analysis

As discussed, the Gouy-Chapman theory of ion distributions assumes that the energy $E_i$ in the Boltzmann factor in Eq.(4) depends solely upon the mean field electrostatic potential. As a result, structural properties of the liquid are ignored, including the ion or solvent sizes and interactions between ions and solvent molecules. These properties lead to packing effects and correlations (ion-solvent, solvent-solvent, and ion-ion) that influence the ion free energy. The liquid structure can be included formally by expressing $E_i(z)$ as

$$E_i(z) = e_i\phi(z) + f_i(z), \tag{8}$$

where $f_i(z)$ is a free energy profile of ion $i$ that describes the correlations [4, 29].

We adopted two different approaches to determine the free energy profile $f_i(z)$. In the first approach, we substitute the potential of mean force (PMF) computed from molecular dynamics (MD) simulations for $f_i(z)$. The MD PMF is determined by calculating the mean force on a single ion positioned at different distances from the interface between pure water and pure nitrobenzene. To calculate the exact $E_i(z)$ in Eq. (4) from MD simulations requires the consideration of ion-ion interactions, but this is not computationally feasible at present. Instead, we approximate $E_i(z)$ by a sum of the electrostatic term $e_i\phi(z)$ and the potential of mean force for a single ion [82].

In the second approach, we postulate a simple functional form for the free energy profile $f_i(z)$, which is referred to as the analytic potential of mean force. This phenomenological function allows the data to be systematically fit to determine the sensitivity of our measurement to features in the potential of mean force.

In both approaches, the ion distributions are determined by solving Eq.(4) numerically with $E_i(z) = e_i\phi(z) + f_i(z)$ and the approximation $\varepsilon(z) \equiv \varepsilon$. Additional analyses will relax the assumption of $\varepsilon(z) \equiv \varepsilon$ and include ion-ion interactions calculated using the hypernetted chain closure for the ion-ion correlation functions.

**3.3.1 Molecular Dynamics Simulations of the Potential of Mean Force**

The potentials of mean force calculated from MD simulations (see Materials and Methods) for TBA$^+$ and Br$^-$ at the nitrobenzene-water interface are shown in Figure 6. Ions can penetrate and transfer through a liquid-liquid interface as illustrated by the continuity across the interface of the potential of mean force. The ion diameter and hydration/solvation effects contribute to the distance required for the potential of mean force to cross from one bulk value to the other. We have not calculated the potential of mean force for TPB$^-$ ions, so we postulate a simple functional form for TPB$^-$ that has these qualitative features (see Figure 6 and Section 3.3.2). Since the interfacial concentration of TPB$^-$ is small, the electron density calculation is not sensitive to this function.

The potentials of mean force for TBA$^+$, TPB$^-$, and Br$^-$ are used for the functions $f_i(z)$ in Eq.(8) to determine the ion distributions in Eq.(5) by solving Eq.(4) for $\phi(z)$ with the numerical quasi-linearization procedure [83, 84]. Figure 7 illustrates these ion distributions calculated for our samples. The ion distributions vary continuously across the interface, in contrast to the discontinuous Gouy-Chapman distributions. Further from the interface, the ion distributions are qualitatively similar to those calculated from the Gouy-Chapman theory (Fig. 3), but differ significantly near the interface. Broadening of the ion distributions at the interface is expected from the finite sizes of the ions and solvent molecules. For TBA$^+$, the enhanced broadening on the water side of the interface is a result of the reduced slope in the potential of mean force in that region, possibly caused by resistance of the ion to lose its hydration shell.

Figure 4 illustrates the electron density profiles $\langle \rho(z) \rangle_{xy}$ calculated from the ion distributions in Fig. 7. As anticipated from the underlying ion distributions, the electron density near the interface is smaller than that predicted by Gouy-Chapman theory.

Figure 5 demonstrates that the reflectivity calculated from the electron density profiles determined from the MD potential of mean force match the measured reflectivities. We emphasize that this is not a fit and that there are no adjustable parameters.

### 3.3.2 Analytic Model for the Potential of Mean Force

We postulate a phenomenological functional form for $f_i(z)$, in spite of the excellent match of the x-ray data with the predictions of the MD potential of mean force, for the following reasons. The MD calculations are time consuming and a simple analytic approximation would be convenient. Fitting the x-ray data to an analytic form can provide insight into the sensitivity of this experiment to probe the function $f_i(z)$.

We model the form of $f_i(z)$ after an error function. This function provides a smooth, monotonic profile through the interface. The error function is the functional form predicted for capillary wave interfacial profiles (Eq.(1)) and is also very similar to the hyperbolic tangent function that arises from van der Waals theories of liquid interfaces [85]. Therefore, it is related to the type of crossover expected at an interface. The parameterization of $f_i(z)$ given in Eq.(9) allows the physically relevant parameters of decay lengths and value of the free energy at the interface to be adjusted. This allows us to test the dependence of the ion distributions and x-ray reflectivity on these features. It will be demonstrated that the best fit of this model to the x-ray reflectivity is in remarkable agreement with the potential of mean force calculated from MD simulations.

The following functional form was used to model $f_i(z)$:

$$f_i^{model}(z) = \begin{cases} (f_i(0) - f_i^w) \cdot erfc[(z - \delta_{i,w})/L_{i,w}]/erfc(-\delta_{i,w}/L_{i,w}) + f_i^w & \text{(water)} \\ (f_i(0) - f_i^{nb}) \cdot erfc[(z - \delta_{i,nb})/L_{i,nb}]/erfc(-\delta_{i,nb}/L_{i,nb}) + f_i^{nb} & \text{(nitrobenzene)} \end{cases} \quad (9)$$

where $f_i(0)$ is the value at $z = 0$, $f_i^a$ is the value for ion $i$ in the bulk phase $a$ (= $w$ for water or $nb$ for nitrobenzene), $f_i^{nb} - f_i^w = \Delta G_{tr}^{o,w \rightarrow nb}$ is the Gibbs energy of transfer of ion $i$ from water to nitrobenzene, the complement of the error function, $erfc(z) = 1 - erf(z)$ (where $erf(z) = (2/\sqrt{\pi})\int_0^z e^{-t^2}dt$ ), $L_{i,a}$ is the decay length that describes the variation of the free energy of ion $i$ from the $f(0)$ value to its value in bulk water or nitrobenzene, and the offset $\delta$ is required to provide continuity of the first and second derivatives of $f(z)$ at $z = 0$. Using these constraints on the derivatives, the known values of $\Delta G_{tr}^{o,w\rightarrow nb}$, and the freedom to set $f_i^w = 0$, only three parameters for each ion are independent. We chose these three parameters to be $L_{i,w}$, $L_{i,nb}$, and $f_i(0)$.

Table 3 lists the best fit parameters when fitting the model in Eq.(9) to our x-ray reflectivity data. Figure 6 shows that the potentials of mean force produced by fitting are similar to the MD simulations. Figure 2 illustrates the fits to the reflectivity data that result from the model in Eq. (9). These fits provide a match to the data of similar accuracy as the reflectivity calculated from the MD potential of mean force (Fig. 5).

Figure 8A compares the ion distributions for the 0.08 M TBABr sample calculated from the MD potential of mean force and from the best fit to the x-ray data using the analytic potential of mean force in Eq. (9). The two sets of distributions are similar except that the small shoulders near the interface in the TBA$^+$ distribution from the MD PMF are absent from the analytic model. The shoulder on the water side of the interface is a result of the reduced slope in the potential of mean force in that region (Fig. 6). Both sets of ion distributions can be used to predict reflectivities that closely match the experimental data, indicating that the x-ray reflectivity measurements are not sensitive to the small differences between them.

Figure 8B illustrates the separate energy terms from the electrostatic potential and the potential of mean force whose sum produces the energy $\Delta E_{TBA^+}(z)$ used in Eq.(5) to calculate the ion distribution for the TBA$^+$ ion. The functional form of these energies is directly related to the form of the ion distributions in Fig. 8A. For example, the shoulder in the TBA$^+$ distribution on the water side is due to the variation in slope of the MD PMF, whereas the shoulder on the nitrobenzene side is due to the relative slopes of the electrostatic potential and the PMF. The shoulders are subtle features not probed by the x-ray measurements.

The decay lengths $L_{i,w}$ and $L_{i,nb}$ should be influenced by the ion sizes, though other effects such hydration and solvation may be important. We investigated whether a simple interface model, that sets the values of the decay lengths equal to the ion diameters and sets $f(0)$ equal to half the Gibbs energy of transfer, provides a good characterization of the ion distributions. Setting the decay lengths $L_{Br^-,w} = L_{Br^-,nb} = 3.7\text{Å}$, $L_{TBA^+,w} = L_{TBA^+,nb} = 8.6\text{Å}$, $L_{TPB^-,w} = L_{TPB^-,nb} = 9.5\text{Å}$, and setting $f(0) = \Delta G_{tr}^{o;w \to nb}/2$ provides an appealing simple model for the potential of mean force. These values for the parameters are within the ±2-sigma error bars shown in Table 3. The reflectivity predicted from this model does not match the reflectivity as well as that predicted by the MD potential of mean force or calculated from the best fit analytic potential of mean force. However, it provides a better match to the data than the Gouy-Chapman, modified Verwey-Niessen, or MPB5 models (the latter two will be discussed below), and may be a useful first approximation under appropriate circumstances. For the 0.08 M TBABr sample, the deviation of this simple model from the data is approximately five standard deviations, as shown in Fig. 9.

### 3.4  Ion-Ion Interactions

We follow Kjellander's formulation to account for ion-ion interactions that include ion-ion correlations and image charge effects [86-88]. This formulation uses the hypernetted chain closure scheme for the ion-ion pair correlation functions. Our application of this approach is outlined in Appendix A. This approach allows us to modify Eq. (8) for the energy of ion $i$,

$$E_i(z) = e_i\phi(z) + f_i(z) + W_i(z), \tag{10}$$

where $W_i(z)$ is the excess chemical potential that accounts for ion-ion correlations that include image forces (see Appendix A), and $f_i(z)$ is the potential of mean force discussed in Section 3.3 that describes the ion-solvent interactions and ion size effects. Figure 10 illustrates $W_i(z)$ near the interface for the 0.08 M TBABr sample. The discontinuities in $W_i(z)$ at $z=0$ are due to the approximation of a constant permittivity function. Calculations with a continuous permittivity function (modeled as an error function interfacial crossover with a width of approximately 5 Å) produce a very similar, though continuous, function $W_i(z)$. However, use of this continuous version of $W_i(z)$ does not alter our conclusion, to be discussed, about the effect of ion-ion correlations in our samples.

Figure 11 illustrates ion distributions given by Eq. (5) that are determined by a numerical solution of Eq.(4) utilizing the expression for $E_i(z)$ in Eq.(10) and the analytical expression for the potential of mean force in Eq.(9). Figure 11 shows that the excess chemical potential $W_i(z)$ has a very small effect on the ion distributions in our samples.

As expected, the reflectivity calculated from the distributions shown in Fig.11 (with the addition of the $TPB^-$ ion distributions previously discussed) is very similar to the reflectivity calculated without $W_i(z)$. This is illustrated in Fig. 9 which shows the difference, in terms of experimental standard deviations (or error bars), between the predicted reflectivity and the measurements for the 0.08 M TBABr sample. The difference between the calculated reflectivity and the measurements is almost indistinguishable for the calculations based upon the molecular dynamics PMF, the analytic PMF, and the analytic PMF plus the excess chemical potential.

### 3.5 Modified Verwey-Niessen Model

The modified Verwey-Niessen model describes the liquid-liquid interface as consisting of a compact inner layer that separates two outer layers of diffuse Gouy-Chapman ion distributions in the two bulk phases [24, 25]. Tension measurements indicate that specific ion adsorption does not occur in our system [25], therefore, the compact inner layer would consist only of solvent (water and nitrobenzene) molecules (see Fig. 12). This form for the ion distributions disagrees with MD simulations which do not provide any evidence for ion-free solvent layers at the interface. However, this model has been discussed extensively in the electrochemical literature (see, for example, [16, 20, 89-91], and references within) and our analysis tested it against our reflectivity data.

To judge the applicability of the modified Verwey-Niessen model to the interpretation of our reflectivity data, we calculated the reflectivity from this model in two different ways. The first consisted of fixing the closest approach of the center of water-phase $Br^-$ ions to the interface to be 4.6 Å and the closest approach of the center of nitrobenzene-phase $TBA^+$ ions to the interface to be 13.3 Å (Fig. 12) This yielded a single layer of pure water on the water-side of the interface and a single layer of pure nitrobenzene on the nitrobenzene-side of the interface. The Gouy-Chapman ion distribution was used at larger distances from the interface. In the second approach we varied the thickness of the pure water and nitrobenzene layers at the interface to produce the best fit to the x-ray data. This led to a closest approach of the center of water-phase $Br^-$ ions to the interface to be 2 Å (i.e., ~ no water layer) and the closest approach of the center of nitrobenzene-phase $TBA^+$ ions to the interface to be 25 Å. In both approaches, Eqs. (4) and (5) are solved as previously discussed, and the predicted ion distributions are converted to electron density profiles (Eqs. (6) and (7)). Both approaches gave similar results. By reducing the maximum ion concentration at the interface this model provides a better match to the x-ray reflectivity data than the Gouy-Chapman model, but still differs from the measurements by approximately eight standard deviations. Figure 9 illustrates the difference, in terms of experimental standard deviations (or error bars), between the predicted reflectivity and the measurements for the 0.08 M TBABr sample.

**3.6 Modified Poisson-Boltzmann (MPB) Theory**

A considerable literature exists on modifications of the Poisson-Boltzmann equation. Kirkwood showed that the approximations in the Debye-Hückel theory consisted of neglecting excluded volume and fluctuation effects [4]. The modified Poisson-Boltzmann (MPB) theories use Leob's closure [92] to estimate the fluctuation term in Kirkwood's theory, and provide a method to calculate the mean electrostatic potential around an ion with spherical exclusion volume (see [5] for a review). Levine, Outhwaite and Bhuiyan developed several versions of the MPB theory, including the latest that we use, MPB5, for a double layer at a hard planar interface [5, 84, 93, 94].

The ion distribution at the interface can be calculated numerically within the MPB5 model [94] by using the quasi-linearization technique [83, 84]. Here, we follow the MPB5 theory for a planar interface, which uses a hard-wall interface model that assumes that ions in one phase cannot move to the other phase. In each phase, the double layer ion distribution can be calculated using Eq.(4) with a constant permittivity function and the following expression for the ion distribution [94]:

$$c_i(z) = c_i^o \zeta_i \exp\{-\tfrac{1}{2kT} e_i[\frac{e_i}{4\pi\varepsilon a}(F - F_0) + F\phi(z+a) + F\phi(z-a) - \frac{F-1}{2}\int_{z-a}^{z+a}\phi(Z)dZ]\} \quad (11)$$

where $a$ is the radius of the ion and the functions $\zeta_i$ and $F$ are given in Ref. [94]. We use the radius of Br$^-$ for the value of $a$ in the water phase, and the radius of TBA$^+$ for the value of $a$ in the nitrobenzene phase. Similar results are obtained when using the average of the ion sizes for the parameter $a$, or when making $a$ slightly larger to include ion hydration.

The ion distributions of Br$^-$ and TBA$^+$ calculated with MPB5 are qualitatively similar to the distributions calculated from the modified Verwey-Niessen theory in the sense that there is a region adjacent to the interface that does not include the center of any ions because the MPB5 theory treats finite sized spherical ions packed against a hard wall. However, the ion density is higher in the MPB5 theory because of ion-ion interactions. The predicted ion distributions (not shown) are converted to electron density profiles as previously discussed (in Eqs. (6) and (7)).

The reflectivity predicted from the MPB5 theory differs from the x-ray measurements by approximately ten standard deviations. Figure 9 illustrates the difference, in terms of experimental standard deviations (or error bars), between the predicted reflectivity and the measurements for the 0.08 M TBABr sample.

## 4. Discussion

These results demonstrate the importance of liquid structure on interfacial ion distributions. The predictions of a mean field theory that neglects liquid structure, the Gouy-Chapman theory, progressively differ from our x-ray measurements of samples with increasing concentration (Figs. 5 and 9). At the highest concentration, the Gouy-Chapman prediction varies from our data by twenty-five standard deviations (Fig. 9). The agreement between the predictions from the potential of mean force calculated by MD simulations and the x-ray measurements indicates that the aspects of liquid structure included in the MD simulations, such as ion sizes and ion-solvent interactions, are required to predict the ion distributions probed by the x-ray measurements.

The parameterization of a phenomenological (analytic) potential of mean force given by Eq. (9) provides insight into the features of the potential of mean force necessary to explain our data. The best fit parameters determined by fitting the x-ray reflectivity to Eq. (9) yield a potential of mean force for $TBA^+$ and $Br^-$ that matches the results from the MD simulation closely, except that the sudden change of slope in the MD simulated PMF for $TBA^+$ is not included in the analytic PMF [95]. Since the analytic PMF provides an excellent fit to the x-ray data, this change of slope is not required to explain our measurements. The fit to the analytic PMF is of equivalent accuracy to the match between the data and the prediction from the MD PMF (Fig. 9).

A simple approximation to the ion distribution can be calculated from Eq. (9) by assigning the decay lengths $L_i$ to be the ion diameters and the crossover free energy $f_i(0)$ to be half the Gibbs energy of transfer. This *ad hoc* simple interface model does not match the x-ray reflectivity as well as the best-fit analytic PMF or the MD PMF. However, this model may provide a useful estimate of the ion distribution under appropriate circumstances.

The MD simulations do not include ion-ion correlations that are expected to be important at high concentrations. The match of the x-ray reflectivity data with the prediction from the MD simulations suggests that these correlations do not significantly affect the ion distributions probed in this experiment, most likely because the correlations are weak at the concentrations of our samples. Calculations that included ion-ion interactions via an HNC closure confirm that ion-ion correlations have a weak effect on ion distributions in our samples (Fig. 9).

The predictions of the modified Verwey-Niessen and MPB5 models were also tested against our data. Except for the most dilute samples the predictions of both models varied significantly from our reflectivity measurements (Fig. 9), though both provided better agreement than the Gouy-Chapman theory. Both models incorporate an ion-free region at the interface (that does not contain the center of any ions), though for different reasons: by postulate for the modified Verwey-Niessen, and as a result of finite sized ions packing against a hard wall for the MPB5 theory. This feature does not agree with the continuous ion distributions that were suggested to be important [37] and that are produced by MD simulations. Our structural data indicates that these models do not adequately describe the ion distributions.

In addition to demonstrating that interfacial liquid structure is required to explain our structural measurements of ion distributions, this work provides a method for including the liquid structure in the analysis of structural measurements of ion distributions near charged or electrified interfaces. This method involves calculating the potentials of mean force by MD simulations or by analytic theory, then using the potential of mean force to predict the ion distributions for a particular experimental situation. We anticipate that this method can be applied to study ion distributions near charged solid surfaces, liquid-vapor interfaces, and the surfaces of charged biomolecules.

**Acknowledgments**

MLS and PV acknowledge support from NSF-CHE0315691, IB from NSF-CHE0345361. MLS thanks Jeff Gebhardt, Tim Graber, and Harold Brewer for help with the ChemMatCARS ID beamline and Binyang Hou for assisting with the x-ray measurements.

ChemMatCARS is supported by NSF-CHE, NSF-DMR, and the DOE. The APS is supported by the DOE Office of Basic Energy Sciences.

**Appendix A**

We outline our application of the formalism developed by Kjellander for ion-ion interactions that account for ion-ion correlations, including image forces, in an inhomogeneous planar system [86-88]. The solvent is treated as a dielectric continuum with a sharp interface between the water and nitrobenzene phases. Kjellander demonstrated an isomorphic mapping from a planar inhomogeneous system subdivided into $M$ layers to a homogeneous 2-dimensional $M$-component mixture. Ions within a layer interact with ions in all layers. Kjellander illustrated the use of the hypernetted chain closure with his formalism to predict ion distributions near planar surfaces.

The hypernetted chain (HNC) closure can be expressed as:

$$c^{ij}(r) = h^{ij}(r) - \ln g^{ij}(r) - \beta u^{ij}(r)$$
$$= h^{ij}(r) - y^{ij}(r) \tag{A1}$$

where the superscripts $i$ and $j$ (i.e., the $z$-coordinates) label the layers for each type of ion, $g^{ij}(r)$ is the radial distribution function, $h^{ij} \equiv g^{ij} - 1$ is the pair correlation function; $c^{ij}$ is the direct correlation function introduced by Orstein and Zernike, $y^{ij} \equiv \ln g^{ij} + \beta u^{ij}$, $r$ is the coordinate in the plane of the layer, $\beta = (kT)^{-1}$, and $u$ is taken as the Coulomb interaction.

A sharp interface model for the liquid-liquid interface employs relative permittivities $\varepsilon_{r,1}$ and $\varepsilon_{r,2}$ that are homogeneous in phase 1 (water) and 2 (nitrobenzene). The Coulomb interaction between ion 1 (charge $q_1$ at a distance $z_1$ from the interface) and ion 2 (charge $q_2$ at a distance $z_2$ from the interface) in the same phase 1 can be expressed as:

$$u(r) = \frac{q_1 \cdot q_2}{4\pi\varepsilon_o \varepsilon_{r,1} [r^2 + (z_1 - z_2)^2]^{1/2}} + \frac{f \cdot q_1 \cdot q_2}{4\pi\varepsilon_o \varepsilon_{r,1} [r^2 + (z_1 + z_2)^2]^{1/2}} \tag{A2}$$

where $f = -\dfrac{\varepsilon_{r,2} - \varepsilon_{r,1}}{\varepsilon_{r,2} + \varepsilon_{r,1}}$ and $r^2 = x^2 + y^2$ is the distance between ions 1 and 2 in a plane parallel to the interface, when the total distance between the two ions is larger than the average diameter of the two ions. The second term in Eq. (A2) is the image force at the interface. The Coulomb interaction between ion 1 in phase 1 and ion 2 in phase 2 can be expressed as:

$$u(r) = \frac{2 \cdot q_1 \cdot q_2}{4\pi\varepsilon_o(\varepsilon_{r,1}+\varepsilon_{r,2})(r^2+z^2)^{1/2}} \quad (A3)$$

when the total distance between the two ions is larger than the average diameter of the two ions. When the two ions are closer than their average diameter, the energy $u(r)$ is taken to be infinite (hard-sphere model).

Given $y^{ij}$, $h^{ij}$ can be obtained from

$$h^{ij}(r) = \exp[y^{ij}(r) - \beta u^{ij}(r)] - 1. \quad (A4)$$

The direct correlation function $c^{ij}$ can be obtained by substituting Eq. (A4) into Eq. (A1). The Ornstein-Zernike equation for the multilayer system can be expressed as

$$h^{ij}(r) = c^{ij}(r) + \sum_{kl} \int h^{ik}(r) n^{kl}(r) c^{lj}(r) \cdot dr \quad (A5)$$

where $n^{kl}(z) = \rho^k(z) \cdot \Delta z^k \cdot \delta_{kl}$, $\rho^k(z)$ is the average density of layer $k$, $\Delta z^k$ is the thickness of layer $k$, and $\delta_{kl}$ is the delta-function.

In Fourier space, the Ornstein-Zernike equation can be expressed as:

$$\hat{h}^{ij}(k) = \hat{c}^{ij}(k) + \sum_{kl} \hat{h}^{ik}(k) n^{kl}(z) \hat{c}^{lj}(k) \quad (A6)$$

where $\hat{\phantom{x}}$ refers to a 2-dimensional Fourier transformation (see Eq. 3.4 in Ref. [87]). A procedure for numerical calculation of the 2-d Fourier transformation can be found in Ref. [96]. From Eq. (A1), $\hat{h}^{ij}(k) = \hat{c}^{ij}(k) + \hat{y}^{ij}(k)$. Substituting into Eq. (A5) yields

$$\hat{y}^{ij}(k) = \sum_{kl} \hat{c}^{ij}(k) n^{kl}(z) \hat{c}^{ij}(k) + \sum_{kl} \hat{y}^{ij}(k) n^{kl}(z) \hat{c}^{ij}(k). \quad (A7)$$

Expressing Eq. (A7) in matrix form,

$$\mathbf{Y(1-NC) = CNC} \quad (A8)$$

Since the Coulomb interaction is a long-range interaction, $g^{ij}(r)$, $c^{ij}(r)$ and $y^{ij}(r)$ are long-range functions. They can be written as a sum of a short-range part and a long-range part to address the difficulty of cut-off errors in the calculation of the Fourier transform over a limited range in $r$ and $k$ [96, 97]:

$$g^{ij}(r) = g_S^{ij}(r) + g_L^{ij}(r) \quad (A9.1)$$

$$c^{ij}(r) = c_S^{ij}(r) + c_L^{ij}(r) \tag{A9.2}$$

$$y^{ij}(r) = y_S^{ij}(r) + y_L^{ij}(r) \tag{A9.3}$$

Eq. (A1) indicates that the direct correlation function has the usual Coulomb limit for large $r$ i.e., $c^{ij}(r) \approx -\beta u^{ij}(r)$ for large $r$ because $g^{ij}(r) \approx 1$ and $h^{ij}(r) \approx 0$ for large $r$ [97]. When $i$ and $j$ refer to the same layer with the Coulomb interaction given by $\frac{q^i q^j}{4\pi\varepsilon_o \varepsilon_r} \cdot \frac{1}{r}$, the long-range piece of the functions in Eq. (A9) and their 2-dimensional Fourier transforms can be chosen as [97]:

$$c_L^{ij}(r) = (-\beta U^{ij}/\alpha) \int_0^\alpha dt (\alpha e^{-rt} - t e^{-r\alpha}) \tag{A10.1}$$

$$y_L^{ij}(r) = (\beta U^{ij} k_D / \alpha) \int_0^\alpha dt (\alpha e^{-rt} - r e^{-r\alpha})(t^2 + k_D^2)^{-1/2} \tag{A10.2}$$

$$\hat{c}_L^{ij}(k) = -2\pi \beta U^{ij} \int_0^\alpha dt \cdot t[(t^2+k^2)^{-3/2} - (\alpha^2+k^2)^{-3/2}] \tag{A10.3}$$

$$\hat{y}_L^{ij}(k) = 2\pi \beta U^{ij} k_D \int_0^\alpha dt (t^2 + k_D^2)^{-1/2} t[(t^2+k^2)^{-3/2} - (\alpha^2+k^2)^{-3/2}] \tag{A10.4}$$

where $U^{ij} = \frac{q^i q^j}{4\pi\varepsilon_0 \varepsilon_r}$, $k_D = 2\pi\beta U^{ij} \rho^j \Delta z^j$, $\alpha \approx 20/R$, and $R$ is the range of $r$ in calculation. Note that in Eq. (31) of reference [97], $c_L$ was incorrectly written with an additional $k_D$ as compared to Eq.(A10.3).

When $i$ and $j$ refer to different layers or to the image interaction, the Coulomb interaction is expressed in the form $\frac{q^i q^j}{4\pi\varepsilon_0 \varepsilon_r} \cdot \frac{1}{(r^2+d^2)^{1/2}}$. In this case $d = |z_1 - z_2|$ when the layers are in same phase; $d = |z_1 + z_2|$ and $\varepsilon_r = (\varepsilon_{r,1} + \varepsilon_{r,2})/2$ when the layers are in different phases; $d = |z_1 + z_2|$ and $\varepsilon_r = \varepsilon_{r,1}/f$ for the image force interaction when the two ions are in the same phase, say phase 1. Since the 2-dimensional Fourier transform of $\frac{1}{(r^2+d^2)^{1/2}}$ $(= 2\pi \exp(-d \cdot k)/k)$ decays fast enough for large $k$, the long-range functions and their 2-dimensional Fourier transforms can be written as [87]:

$$c_L^{ij}(r) = -\beta u^{ij}(r) \tag{A11.1}$$

$$y_L^{ij}(r) = \beta u^{ij}(r) \tag{A11.2}$$

$$\hat{c}_L^{ij}(k) = -2\pi\beta \hat{u}^{ij}(k) \tag{A11.3}$$

$$\hat{y}_L^{ij}(k) = 2\pi\beta \hat{u}^{ij}(k) \tag{A11.4}$$

A further difficulty arises from the discontinuity in $h^{ij}(r)$ and $c^{ij}(r)$ in the hard-core sphere model, when $d < a$ ($a$ is the average diameter of the two ions). The discontinuity induces a long range tail in $k$ space. Kjellander introduced the function $Q^{ij}(r)$ defined within the hard-core region $r < (a^2 - d^2)^{1/2}$ to compensate for the discontinuity in $h^{ij}(r)$ and its derivative $h'^{ij}(r)$ [87],

$$Q^{ij}(r) = \begin{cases} \Delta h^{ij} + (r^2 - a^2)\Delta h'^{ij}/2a, & r < (a^2 - d^2)^{1/2} \\ 0, & r > (a^2 - d^2)^{1/2} \end{cases} \tag{A12}$$

where $\Delta h^{ij}$ and $\Delta h'^{ij}$ are the discontinuity of $h^{ij}(r)$ and its derivative at $(r^2 + d^2)^{1/2} = a$. The Fourier transform of $Q^{ij}(r)$ is given as:

$$\hat{Q}^{ij}(k) = 2\pi a[\Delta h^{ij} J_1(ak)/k - \Delta h'^{ij} J_2(ak)/k^2] \tag{A13}$$

where $J_1$ and $J_2$ are integer order Bessel functions. Since $y^{ij}(r)$ is continuous in the whole space, Eq. (A12) compensates for the discontinuity in $c^{ij}(r)$ at $r^2 + d^2 = a^2$ and Eq. (A13) can be used in the calculation of $\hat{c}^{ij}(k)$.

The following functions are defined for the Fourier transformations of the short range functions:

$$\tilde{c}_S(r) = c_S(r) + Q(r) \tag{A14.1}$$

$$\tilde{h}_S(r) = h_S(r) + Q(r) \tag{A14.2}$$

$$\hat{\tilde{c}}_S(k) = \hat{c}_S(k) + \hat{Q}(k) \tag{A14.3}$$

$$\hat{\tilde{h}}_S(k) = \hat{h}_S(k) + \hat{Q}(k) \tag{A14.4}$$

The short-range functions can be obtained from the procedure in Scheme A1. When the ion concentration is low, $y^{ij}(r) = 0$ can be used for the initial $y^{ij}$. When the concentration is high, say $n^{kl}(z) > 0.3$ M, $y^{ij}(r) = 0$ is not a good initial function. In that case, $y^{ij}$ determined for a lower concentration can be used as the initial $y^{ij}$.

The excess chemical potential $W^i$ that accounts for ion-ion correlations that include image forces can be calculated using

$$W^i = \frac{1}{\beta}\sum_j \int dr \cdot n^{ij}\{\frac{1}{2}[h^{ij}(r)]^2 - c^{ij}(r)\} + \frac{y^{ii}(0)}{2\beta}. \quad (A15)$$

$W^i$ contains information on the $z$-dependence for all the ions: Br$^-$, TBA$^+$, and TPB$^-$. Using the convention established in the main body of this paper that labels the ions by a subscript $i$, we denote the excess chemical potential as $W_i(z)$. The ion distribution is determined by the Boltzmann equation, similar to Eq.(5):

$$c_i(z) = c_i^o \cdot \exp\{-[e_i\phi(z) + f_i(z) + W_i(z)]/kT\}. \quad (A16)$$

The self-consistent ion distribution was obtained by iteration of the Poisson-Boltzmann equation and the HNC closure calculation as described in Scheme A2.

In Eq. (A15), $W^i$ is calculated by integrating $h^{ij}(r)$ and $c^{ij}(r)$ in real space. To achieve an accurate numerical result, $R$ and the number of points in $r$ space cannot be too small, even though a convergent $h^{ij}(r)$ and $c^{ij}(r)$ can be obtained in the HNC calculation for small values of $R$ and number of points. The choice of these two parameters can be estimated from the calculation of $W^i$ in a homogeneous bulk phase using the 3-dimensional HNC calculation [98, 99].

In our samples, the ion density increases rapidly at the interface region. The layer thickness for the calculations was chosen to be ~1.2 Å in the region where the ion density increases rapidly and 2 to 4 Å in the region away from the interface. Since ion-ion correlations occur primarily within a Debye length of the ion, the number of layers can be limited to produce a sufficiently accurate excess chemical potential (i.e., with an accuracy of 0.01 kJ/mol) for the ion density calculation. The calculation time for the HNC closure increases as $N^2$ (where $N$ is the number of layers times the number of ionic species). To reduce the calculation time, the HNC closure was applied to the ions in each phase by considering only ions in the neighboring phase within several Debye lengths from the interface.

# References

*Authors to whom correspondence should be addressed, E-mail: schloss@uic.edu, pvanysek@niu.edu.

**Tables**

Table 1. Sample properties.

$C_{initial}$ is the molar concentration of the water solution of TBABr before equilibration with a 0.01M TBATPB solution in nitrobenzene (NB). The equilibrated molar concentration of ion $i$ in the two bulk phases is given by $c_i^o$. The volume ratio of the sample is 2 : 1 (water:nitrobenzene). The interfacial electric potential $\Delta\phi^{w-nb}$ (potential in the bulk water phase minus that in the bulk nitrobenzene phase) is given in volts.

| $C_{initial}$ /mol l$^{-1}$ ($\pm 1\%$) | $c_{Br^-}^o$ /mol l$^{-1}$ in water | $c_{TBA^+}^o$ /mol l$^{-1}$ in water | $c_{TPB^-}^o$ /mol l$^{-1}$ in water | $c_{Br^-}^o$ /mol l$^{-1}$ in NB | $c_{TBA^+}^o$ /mol l$^{-1}$ in NB | $c_{TPB^-}^o$ /mol l$^{-1}$ in NB | $\Delta\phi^{w-nb}$ /V |
|---|---|---|---|---|---|---|---|
| 0.010 | 0.00976 | 0.00976 | 3.35x10$^{-13}$ | 0.0005 | 0.0105 | 0.010 | -0.245 |
| 0.040 | 0.0375 | 0.0375 | 1.25x10$^{-13}$ | 0.0050 | 0.0150 | 0.010 | -0.270 |
| 0.050 | 0.0466 | 0.0466 | 1.13x10$^{-13}$ | 0.0069 | 0.0169 | 0.010 | -0.273 |
| 0.057 | 0.0521 | 0.0521 | 1.08x10$^{-13}$ | 0.0081 | 0.0181 | 0.010 | -0.274 |
| 0.080 | 0.0736 | 0.0736 | 9.63x10$^{-14}$ | 0.0128 | 0.0228 | 0.010 | -0.277 |

Table 2. Interfacial tension $\gamma$ and interfacial width (roughness) of the samples. Samples labeled by $C_{initial}$, the initial concentration of TBABr in water. The interfacial width $\sigma$ is determined by fitting the data in Figure 2 to Eq.(2), and the width $\sigma_{cap}$ is calculated from capillary wave theory using Eq.(3) and the tension $\gamma$.

| $C_{initial}$/mol l$^{-1}$ | $10^3$ $\gamma$/ N m$^{-1}$ (±0.2) | $\sigma$/Å (±0.1) | $\sigma_{cap}$/Å (±0.02) |
|---|---|---|---|
| 0.010 | 19.6 | 6.2 | 5.8 |
| 0.040 | 15.7 | 7.2 | 6.5 |
| 0.050 | 14.8 | 7.7 | 6.7 |
| 0.057 | 14.7 | 7.8 | 6.8 |
| 0.080 | 13.0 | 8.0 | 7.2 |

Table 3. Analytic Model Fit Parameters.

Best fit parameters for the analytic model of the potential of mean force (Eq. (9)) determined by fits to the x-ray reflectivity data. Errors are determined by the usual procedure of analyzing the $\chi^2$-space. Errors are not given for the TPB⁻ parameters because the fitting is not sensitive to this function since the concentration of TPB⁻ is very small at the interface. Errors are quoted to ±2 standard deviations (±2-sigma).

| Ion $i$ | $L_{i,w}$/Å (±2-sigma) | $L_{i,nb}$/Å (±2-sigma) | $f_i(0)$/kJ mol$^{-1}$ (±2-sigma) |
|---|---|---|---|
| TBA⁺ | 13 (+7/-5) | 13 (+7/-9) | 11 (+2/-6) |
| Br⁻ | 2 (+8/-1) | 7 (+13/-7) | 14 (+6/-8) |
| TPB⁻ | 10 | 10 | 18.6 |

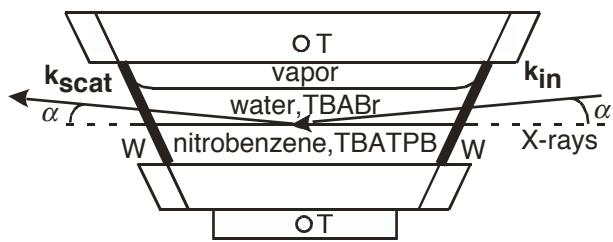

Figure 1  Cross-sectional view of stainless steel sample cell. W - mylar windows; T - thermistors to measure temperature.  The kinematics of surface X-ray reflectivity is also indicated: $k_{in}$ is the incoming X-ray wave vector, $k_{scat}$ is the scattered (reflected) wave vector, $\alpha$ is the angle of incidence and reflection.

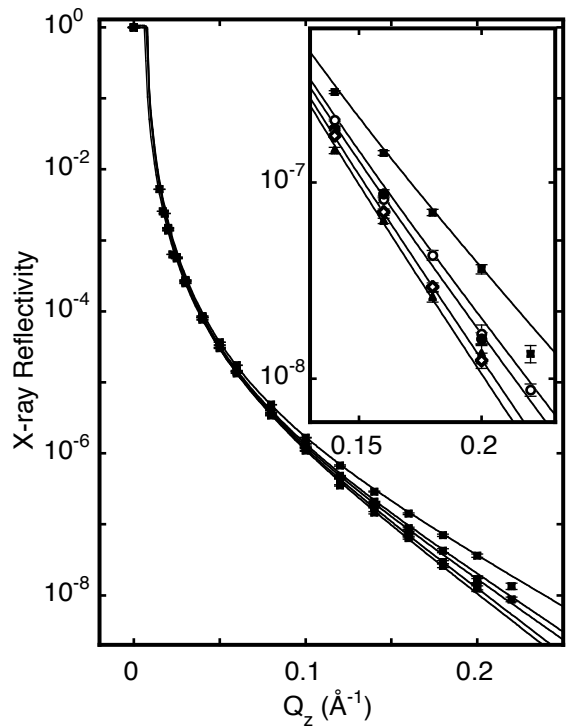

Figure 2. X-ray reflectivity, $R(Q_z)$, as a function of wave vector transfer $Q_z$ from the interface between a 0.01 M solution of TBATPB in nitrobenzene and a solution of TBABr in water at five concentrations (0.01, 0.04, 0.05, 0.057, and 0.08 M top to bottom). Solid lines are predictions using the analytic potential of mean force and are very similar to fits to the error function profile in Eq.(1). Inset shows the data at high $Q_z$ magnified with different symbols for different concentrations: 0.01 M – filled square, 0.04 M – open circle, 0.05 M – filled circle, 0.057 M – open diamond, 0.08 M – filled triangle. Error bars are determined by counting statistics of the x-ray measurements and represent +/- one standard deviation.

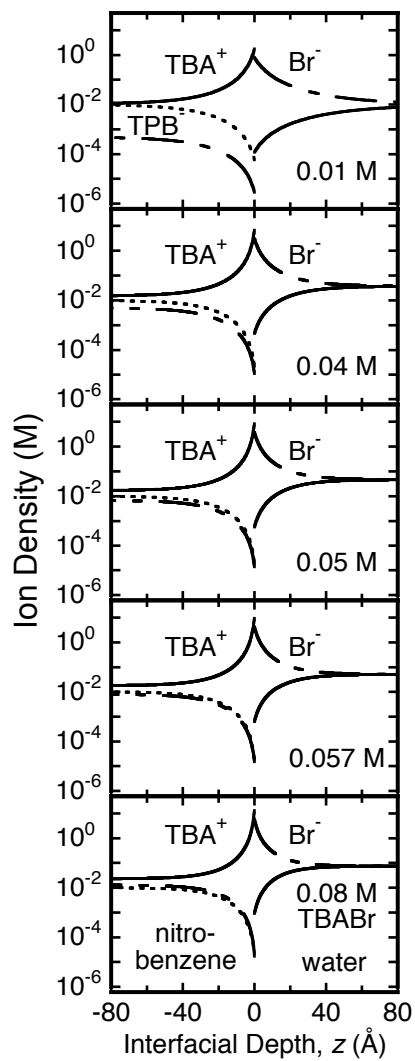

Figure 3. Gouy-Chapman solution for the ion distributions at the interface between a 0.01M TBATPB solution in nitrobenzene ($z < 0$) and a 0.01M (0.04 M, 0.05 M, 0.057 M, 0.08 M) TBABr solution in water ($z > 0$). TBA$^+$ – solid lines, Br$^-$ – short-long dashed lines, TPB$^-$ – short dashed lines.

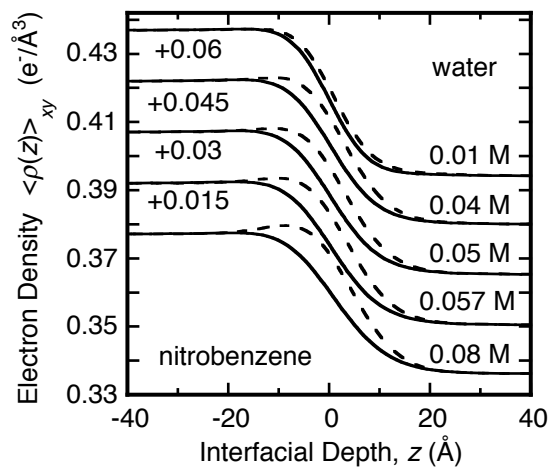

Figure 4. Electron density $\langle\rho(z)\rangle_{xy}$ as function of depth through the interface between a 0.01 M TBATPB solution in nitrobenzene ($z < 0$) and a 0.01 M (0.04 M, 0.05 M, 0.057 M, 0.08 M) TBABr solution in water ($z > 0$). Dashed lines, calculation from Gouy-Chapman model; solid lines, calculation from molecular dynamics simulation of the potential of mean force. The concentrations have been progressively offset by 0.015 along the electron density axis.

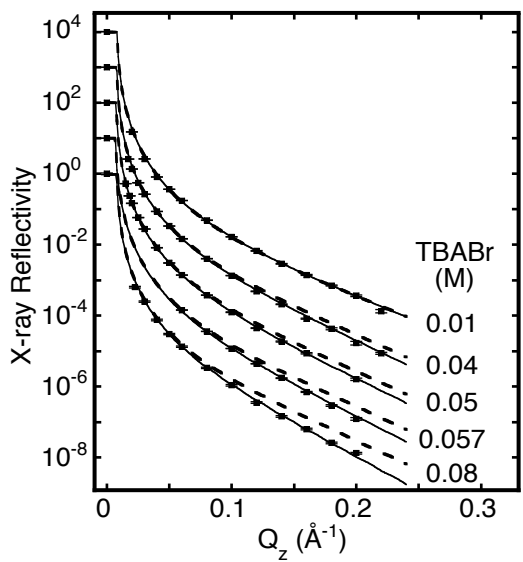

Figure 5. X-ray reflectivity, $R(Q_z)$, as a function of wave vector transfer $Q_z$ from the interface between a 0.01 M solution of TBATPB in nitrobenzene and a solution of TBABr in water at five concentrations (0.01, 0.04, 0.05, 0.057, and 0.08 M, top to bottom, progressively offset by factors of 10 ($R = 1$ at $Q_z = 0$)) at a room temperature of 24 ± 0.5°C. Solid lines are predictions using the potential of mean force from MD simulations. Dashed lines are predicted by the Gouy-Chapman model. No parameters have been adjusted in these two models. Error bars are indicated by horizontal lines through the square data points and are usually much smaller than the size of the squares. The points at $Q_z = 0$ are measured from transmission through the bulk aqueous phase.

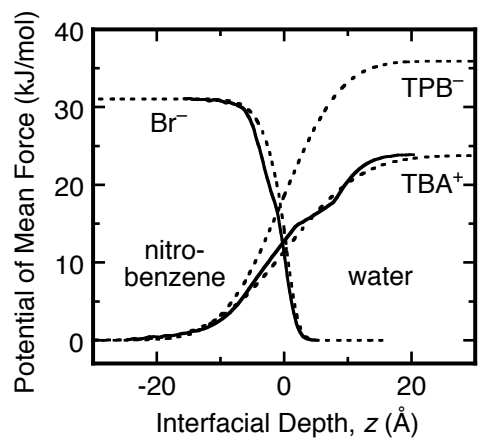

Figure 6. Potential of mean force for $TBA^+$, $Br^-$, and $TPB^-$ at the nitrobenzene-water interface (water is at $z > 0$ and nitrobenzene is at $z < 0$). Solid lines are calculated from MD simulations. Dashed lines are the best fit of the analytic model in Eq. (9) to the reflectivity data (see Table 3 for best fit parameters).

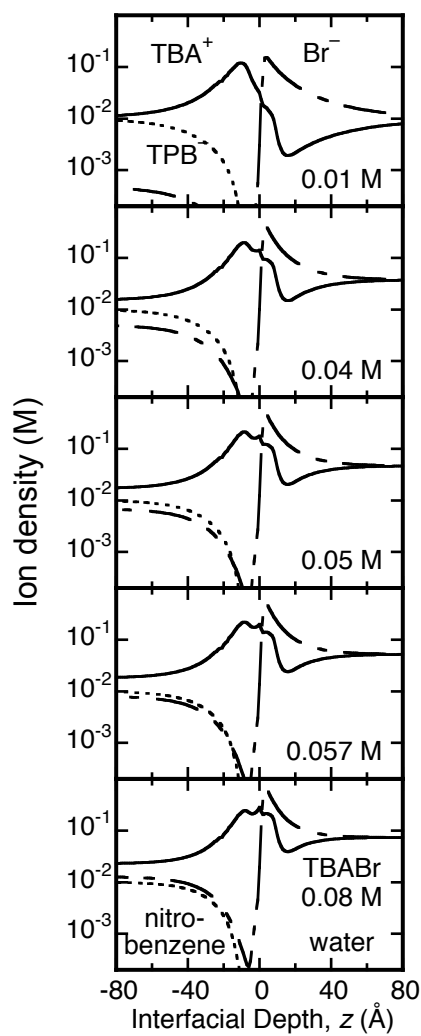

Figure 7. Ion distributions at the interface, calculated from the MD simulation of the potential of mean force, between a 0.01 M, 0.04 M, 0.05 M, 0.057 M, 0.08M (top to bottom) TBABr solution in water ($z > 0$) and a 0.01M TBATPB solution in nitrobenzene ($z < 0$). TBA$^+$ – solid lines, Br$^-$ – short-long dashed lines, TPB$^-$ – short dashed lines.

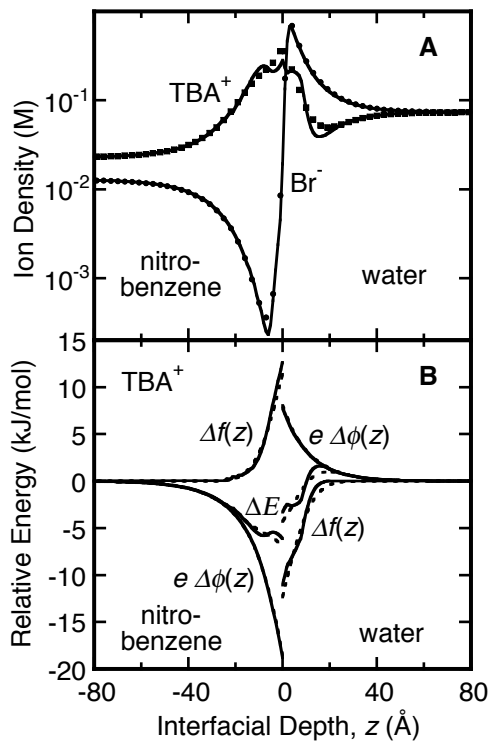

Figure 8. (A) Ion distributions at the interface between a 0.08M TBABr solution in water ($z > 0$) and a 0.01M TBATPB solution in nitrobenzene ($z < 0$), calculated from the MD simulation of the potential of mean force (solid line) and from the analytic model in Eq.(9) using parameters (Table 3) from the best fit to the x-ray data (Br⁻ represented by dots, TBA⁺ by squares). (B) $\Delta E_i(z) = e_i \Delta\phi(z) + \Delta f_i(z)$, that are used to calculate the ion distribution for TBA⁺ in panel (A) (note the similarity to Eq.(8) except that in this figure the energy terms have been subtracted from their bulk values to yield $\Delta E(z)$, $\Delta\phi(z)$, and $\Delta f_i(z)$, rather than $E(z)$, $\phi(z)$, and $f_i(z)$). Solid lines are calculations from the MD potential of mean force and dashed lines are from the best fit of the data to the analytic model in Eq.(9).

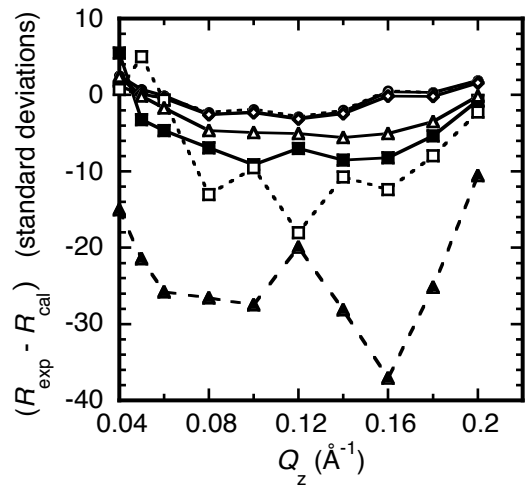

Figure 9  Difference of the calculated reflectivity from the experimental reflectivity (for the 0.08 M TBABr sample) in units of standard deviations (as determined by the experimental statistical error) as a function of wave vector transfer $Q_z$ for the following models.  MD potential of mean force (solid line with dots), best-fit analytic potential of mean force (dashed line with circles), HNC used with the analytic potential of mean force (solid line with hollow diamonds), analytic potential of mean force with decay lengths set to the ion diameters (solid line with hollow triangles), modified Verwey-Niessen (solid line with filled squares), MPB5 (dashed line with hollow squares), and Gouy-Chapman (long dashed line with filled triangles).  The MD, analytic, and HNC-analytic models are nearly coincident.

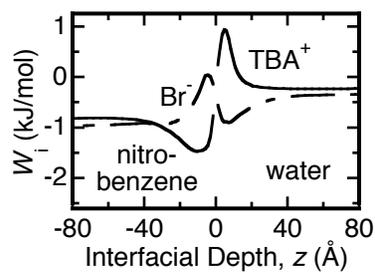

Figure 10  Excess chemical potential $W_i(z)$, that accounts for on-ion correlations that include image forces, calculated for the 0.08 M TBABr sample. $TBA^+$, solid line; $Br^-$, dashed line (water is at $z > 0$ and nitrobenzene is at $z < 0$).

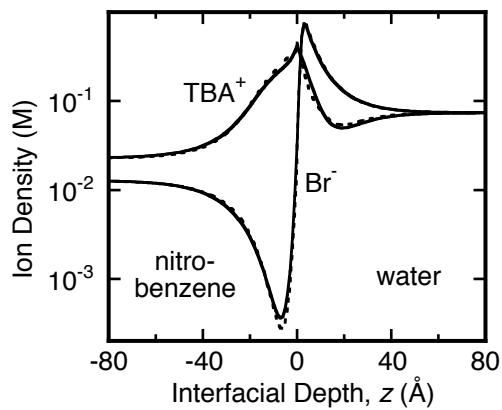

Figure 11 Calculated ion distributions at the interface between a 0.08M TBABr solution in water ($z > 0$) and a 0.01M TBATPB solution in nitrobenzene ($z < 0$): analytic model in Eq.(9) using parameters (Table 3) from the best fit to the x-ray data (solid lines) and analytic model supplemented by ion-ion interactions as in Eq.(10) (dashed lines).

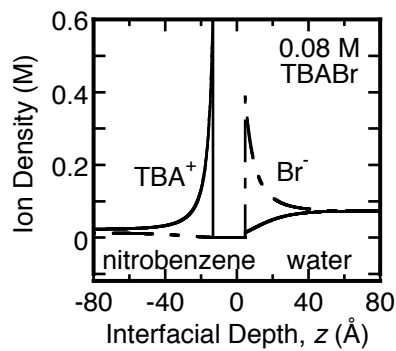

Figure 12  Modified Verwey-Niessen solution for the ion distributions at the interface between a 0.01M TBATPB solution in nitrobenzene ($z < 0$) and a 0.08M TBABr solution in water ($z > 0$).  The closest approach of the center of water-phase Br$^-$ ions to the interface is set at 4.6 Å and the closest approach of the center of nitrobenzene-phase TBA$^+$ ions to the interface is set at 13.3 Å (Br$^-$ – short-long dashed lines, TBA$^+$ – solid lines).  These closest-approach distances model a single layer of ion-free water and ion-free nitrobenzene on either side of the interface.

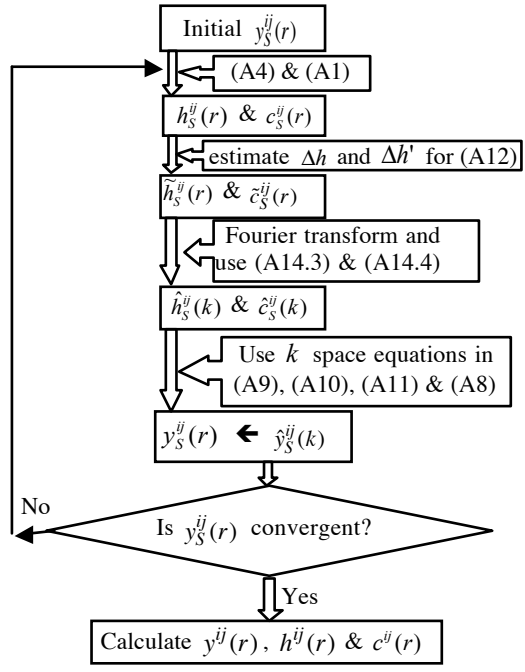

Scheme A1 Iteration procedure for HNC calculation.

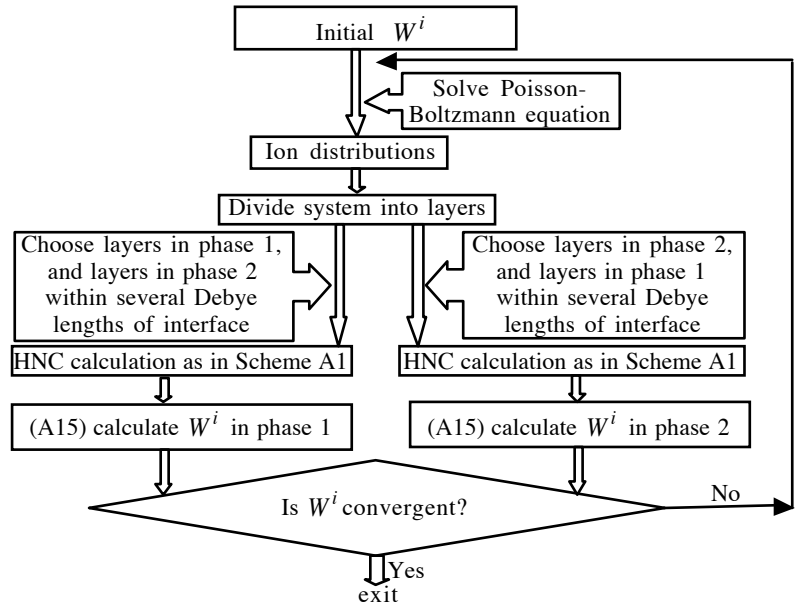

Scheme A2  Iteration procedure that uses Poisson-Boltzmann and HNC calculations to determine a self-consistent ion distribution.